\documentclass[preprint2]{aastex}
\usepackage[]{natbib}
\begin{document}

\title{Further Evidence for Large Intrinsic Redshifts.}

\author{M.B. Bell\altaffilmark{1}}

\altaffiltext{1}{Herzberg Institute of Astrophysics,
National
Research Council of Canada, 100 Sussex Drive, Ottawa,
ON, Canada K1A 0R6}

\begin{abstract}
An examination of the positions and redshifts of the compact, QSO-like objects reported near the Seyfert 2 galaxy NGC 1068 has revealed many relationships that suggest strongly that the objects have been ejected from the galaxy. Although some of the same relations have been found for the compact objects near NGC 3516, the higher number of sources near NGC 1068 has resulted in a detailed physical model. The results indicate that the objects have been expelled in at least four similarly structured ejection events from a point near the centre of NGC 1068 within the last 6$\times10^{6}$ yrs and with very modest velocities [l-o-s components $\sim(1-2)\times10^{4}$ km s$^{-1}$]. From the age and rotation angle of successive events, the rotation period of the central object is calculated to be $\sim10^{7}$ yrs, in good agreement with the nuclear rotation period obtained from kinematic studies of NGC 1068. The large redshifts of the objects cannot be explained by their modest ejection velocities and require an additional, large intrinsic component. 

\end{abstract}


\newpage

\section{Introduction}

Extensive optical and X-ray searches near the Seyfert 2 galaxy NGC 1068 have shown that there are at least 14 compact objects within 50$\arcmin$ of the galaxy ). Eleven of these QSOs-like objects have measured redshifts ranging from 0.261 to 2.018, all significantly higher than the systemic redshift of the galaxy (z = 0.0038, \citet{kra00}). From this high source density it was argued that the objects must be physically associated with NGC 1068 and probably ejected from it \citep{bur99}. 

Evidence for the ejection of objects from a central source has been claimed previously when two objects are found to lie on opposite sides of, and to be equally spaced from, the central object \citep{arp97a,arp97b,bur95,bur97a}. The simplest interpretation may then be that the objects have similar masses and were ejected in the same event. 
It is also well known that matter is ejected from the nucleus of active galaxies and radio galaxies in jets directed along the rotation axis. In several previous cases evidence has been presented that suggests compact objects have also been ejected along the rotation axis of the parent galaxy \citep{arp97a,arp98,bur95, bur97a,bur97b,bur96,chu98,pie94,rad97}. In NGC 1068 the rotation axis, and inner jet, can be represented by a north-south line passing through the galaxy \citep{ant94,gal96,gre96}. 

In most of the above cases the redshifts of the compact objects have been found to decrease with increasing distance from the parent object. 

This paper first examines the source position and redshift data for the fourteen QSOs near NGC 1068 in an attempt to determine if an ejection model can explain them. A possible ejection model is then proposed and further examination leads to ejection velocities and event ages. 

\begin{deluxetable}{cccc}
\tabletypesize{\scriptsize}
\tablecaption{QSOs and BSOs near NGC 1068. \label{tbl-1}}
\tablewidth{0pt}
\tablehead{
\colhead{Object} &  \colhead{Redshift}   &  \colhead{Ang. Dist.}  & \colhead{Mag.}  
\\
         & &  \colhead{from NGC 1068 ($\arcmin)$} 
}

\startdata
1    &  0.261  &  49.4  &    15.7$_{E}$     \\   
2    &  0.468  &  32.9  &    18.8$_{V}$     \\
3    &  0.726  &  48.3  &    18.5$_{V}$     \\
4    &  0.649  &  15.5  &    18.5$_{V}$     \\
5    &  1.054  &  37.2  &    18.0$_{V}$     \\
6    &  1.552  &  13.3  &    18.5$_{V}$     \\
7    &  1.112  &   9.9  &    18.1$_{E}$     \\
8    &  0.385  &   8.4  &    18.4$_{E}$     \\
9    &  -----  &  22.6  &    19.9$_{E}$     \\
10   &  -----  &  6.0   &    19.9$_{E}$     \\
11   &  2.018  &  6.2   &    18.8$_{V}$     \\
12   &  -----  &  12.1  &    19.3$_{E}$     \\
13   &  0.684  &  30.4  &    18.2$_{V}$     \\
14   &  0.655  &  43.9  &    18.2$_{E}$     \\  

\enddata
\end{deluxetable}

\section{Source position and redshift data}

The redshifts and magnitudes of the compact objects, as well as their angular distances from NGC 1068, are listed in Table 1 and have been taken from \citet{bur99}. Source numbers have been retained. 

\section{Examination of the redshift and position data for the QSOs near NGC 1068}

In Fig. 1 the measured redshift of each object is plotted vs its angular separation from NGC 1068. It is apparent from this plot that there are at least four separate groups of sources whose component objects lie at roughly the same distance from the galaxy. The four source groups fall within the boxes indicated in the figure (labeled A, B, C, and D) and each contains three sources. Since redshifts have not been measured for sources 9,10, and 12, their angular separations from NGC 1068 have been indicated only by vertical dashed lines in Fig. 1. When their redshifts are measured, these objects must fall somewhere on their respective dashed lines. As will be seen below, the sources in each group lie close to a circle centered on NGC 1068. If the objects are at cosmological distances, it makes no sense that they should be located on rings around NGC 1068. On the other hand, if the objects have been ejected from NGC 1068 in separate events this is exactly what one might expect. Fig. 1 is therefore interpreted here as indicating that at least four groups of sources have been ejected from NGC 1068 in separate ejection events. 

\begin{figure}
\hspace{-1.0cm}
\vspace{-1.5cm}
\epsscale{1.1}
\plotone{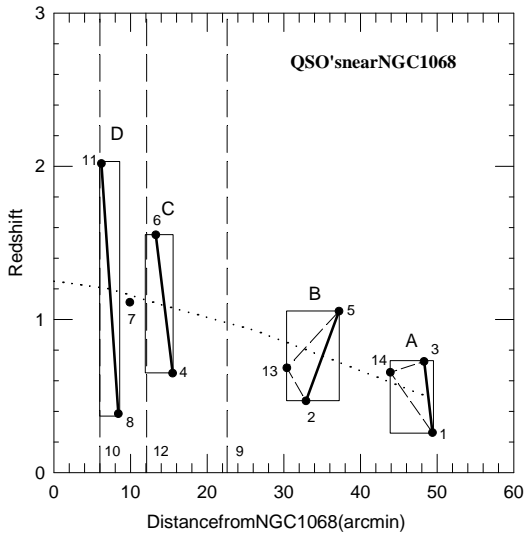}
\caption{\scriptsize{Plot of source redshift vs angular distance from NGC 1068. Boxes A, B, C, and D contain sources that all lie at approximately the same distance form the galaxy. Numbers identify sources and those with no measured redshift are plotted as a vertical dashed line. Heavy solid lines join paired sources whose mean redshift falls off smoothly with increasing distance from NGC 1068.\label{fig1}}}
\end{figure}

The sources ejected in each event are defined unambiguously by the boxes in Fig. 1. The triplet lying furthest from NGC 1068 (box A) is assumed to have been ejected first with close-in triplets being ejected in subsequent events. What follows is an interpretation of the data within the framework of the ejection model defined by Fig. 1. 
 
\section{Additional characteristics of the ejection model as defined by the redshift and position data}

Since the three sources making up each of the triplets defined by the boxes in Fig. 1 all lie at approximately the same distance from NGC 1068, their component sources must fall close to a circle centered near the galaxy. In Fig. 2 the position of each source \citep{bur99} is plotted relative to the position R.A.$_{2000}=02^{\rm h} 42\fm0$; DEC$_{2000}=00\arcdeg00\farcm00$. This choice of reference position is arbitrary and was only chosen to simplify calculations. The location of NGC 1068 is shown by an open square, and the rotation axis is indicated by the vertical dashed line. Four concentric circles, labeled A, B, C, and D as in Fig. 1, have been superimposed on the figure. Although the sources in each triplet fall close to the circles, the fit is poor, especially in the larger triplets. However, if the objects in each triplet all lay at the same radial distance from NGC 1068, they would only fall on concentric circles if they were all ejected in the plane of the sky, perpendicular to the line-of-sight (l-o-s). It is highly unlikely that this would be the case. It is more likely that the ejection planes would be oriented either at different angles to the l-o-s, or at least at some angle other than 90$\arcdeg$. If so, they would fall on concentric ellipses.

\begin{figure}
\hspace{-2.0cm}
\vspace{-2.0cm}
\epsscale{1.05}
\plotone{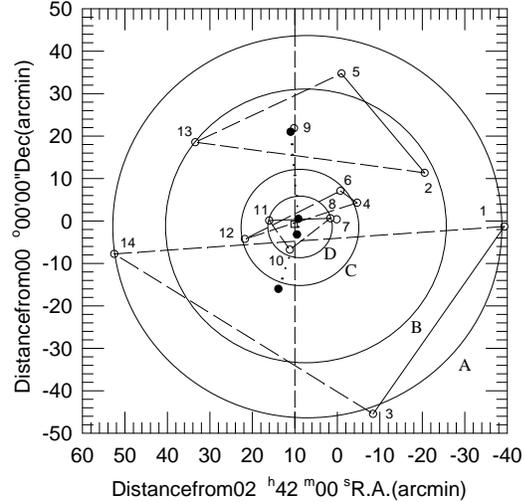}
\caption{\scriptsize{Location of QSO-like objects near NGC 1068. North is at the top, East at the left. Source numbers are from Table 1. Four concentric circles labeled A to D and centered near NGC 1068 have been drawn through the objects identified by the boxes in Fig. 1. The location of NGC 1068 is indicated by the open square. The rotation axis of the central torus is represented by the vertical dashed line through the galaxy.\label{fig2}}}
\end{figure}

In Fig. 3, four concentric ellipses centered near NGC 1068 have been superimposed on the data in Fig. 2. The fit is now excellent, even in the large triplets. The sources making up each triplet, defined by the boxes in Fig. 1, are A(1,3,14), B(2,5,13), C(4,6,12), and D(8,11,10) where source numbers have been taken from Table 1. Sources 7 and 9 have not been fitted to a concentric ellipse but the possibility that one, or both, may be part of a fifth triplet cannot be ruled out.

\begin{figure}
\hspace{-1.5cm}
\vspace{-2.0cm}
\epsscale{1.05}
\plotone{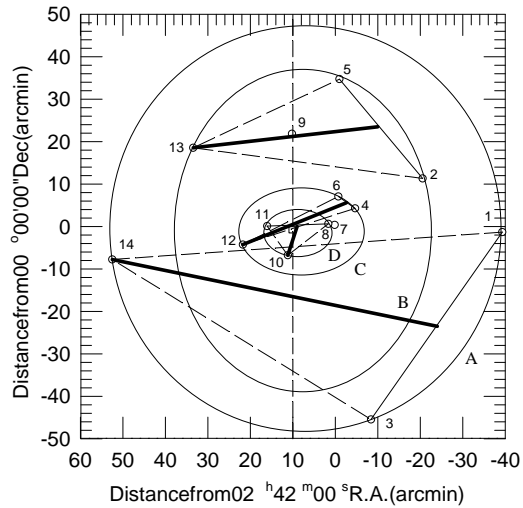}
\caption{\scriptsize{Same as Fig.2 except here ellipses have been fitted to the four triplets identified in Fig. 1. Singlet and pair midpoints are joined by heavy solid lines that show a continuous angular rotation on the sky from triplets A to D. The midpoints of these lines fall close to the rotation axis.\label{fig3}}}
\end{figure}

Also evident in Fig. 3 is the fact that in each triplet, one object (hereafter referred to as the singlet) lies on the East of the rotation axis while the midpoint of the remaining two (hereafter referred to as the pair) is located a similar distance to the West. Pairs are shown joined by a solid line and the relevant singlet is connected to each pair by dashed lines. Heavy solid lines also identify pairs in Fig. 1. It is worth noting that the mean redshift of the pairs falls off approximately linearly with increasing distance from the Seyfert galaxy. This fall-off with increasing distance, indicated by the dotted line in Fig. 1, is similar to what has been found for compact objects near other active galaxies. In Fig. 3 the source positions on the sky are too close to allow an unambiguous choice for the singlet in triplet D (it could be source 10 or 11). It is obvious from Fig. 1, however, that the mean redshift of sources 11 and 8 falls on the dotted line and these two sources are therefore assumed to define the pair. Whether a redshift fall-off with distance is also true for the singlets cannot yet be determined since redshifts have not been measured for source 10 and 12.

The heavy solid line connecting from each singlet to each pair midpoint in Fig. 3 gives the triplet's position angle $\beta$, (where $\beta$ is measured counterclockwise on the sky, or East, from North). The midpoints of these heavy lines, (indicated by the dotted curve in Fig. 2), are assumed to represent the center of mass points of the triplets (simply because the singlets and pair midpoints are equally spaced on either side of the rotation axis). They fall along a curve that is closely aligned with the rotation axis, the minor axis of the central torus \citep{ant94,gal96,gre96} and the inner radio jet \citep{gal96} and counter-jet. Although the center of mass points do not coincide exactly with the rotation axis, their departure from it varies smoothly from triplet A to D and this departure can be explained if NGC 1068 itself has some proper motion perpendicular to the l-o-s over the time period covered by the ejection events. (This is discussed in detail in a later paper). 

The N-S offset between the galaxy and each triplet's center of mass, whether in the jet or counter-jet direction, is assumed to indicate that component of the triplet's motion, resulting from its respective ejection event, that is directed along the rotation axis and perpendicular to the l-o-s. In Fig. 4 the sources in each triplet have been re-plotted after removing the N-S offsets of their respective centers of mass. The plot shows the relative triplet orientation more clearly. Each triplet is identified by an A, B, C, or D adjacent to its associated pair. The triplet position angle $\beta$ and related rotation angle $\theta$ (defined below) rotate smoothly about the rotation axis in a counter-clockwise direction from A to D (again this continuous rotation for triplets A to D is achieved only if source 10 is chosen for the singlet in triplet D). This rotation direction is the same as the galaxy rotation direction and may suggest that the triplet position angle is fixed relative to the co-ordinates of the central object. The change in triplet position angle is thus a measure of the rotation of the central object. This possibility is used below to estimate the rotation period of the central object. 

\begin{figure*}
\hspace{-2.5cm}
\vspace{-2.0cm}
\epsscale{1.15}
\plotone{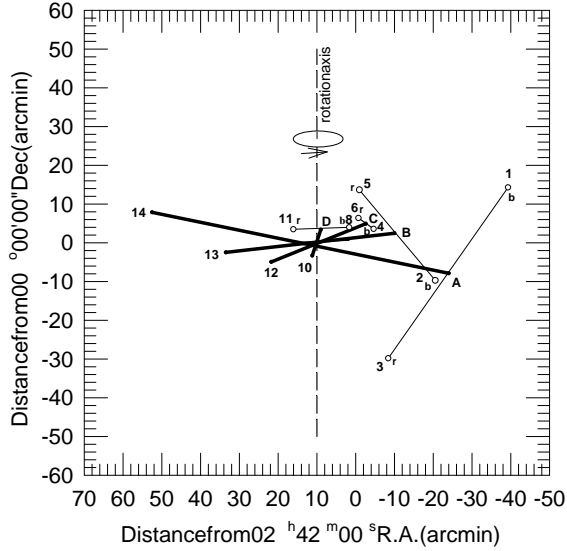}
\caption{\scriptsize{Sources plotted after removing N-S offsets equal to the displacements of their associated triplet mass centers. The letters A, B, C,and D identify the triplets. The "r" or "b" next to the paired sources indicates whether it is redshifted or blueshifted relative to the mean pair redshift. The direction of rotation of the triplet axes (heavy solid lines) is indicated at the top.\label{fig4}}}
\end{figure*}

\begin{figure}
\hspace{-1.5cm}
\vspace{-1.5cm}
\epsscale{1.05}
\plotone{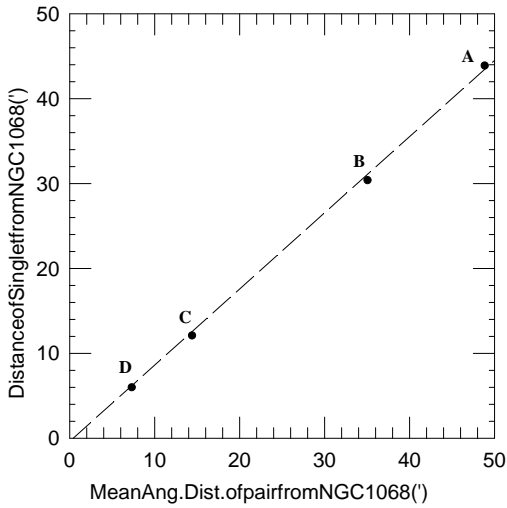}
\caption{\scriptsize{Angular distance of the singlets plotted vs the mean angular distance of the pairs for the four triplets.\label{fig5}}}
\end{figure}

In all four triplets in Fig. 4, the pair orientation angle $\omega$ (defined by the position angle on the sky of the line connecting between the two sources in each pair) rotates continuously with the triplet position angle $\beta$ and, in so doing, maintains a similar orientation relative to the singlet/pair midpoint/rotation axis plane. In effect, all four triplets are similarly structured, differing only in their rotation angles and sizes. To obtain this result by chance seems extremely unlikely if it is remembered that the sources in each triplet were identified in Fig. 1 simply because they lay at similar distances from NGC 1068, and that the internal consistency achieved here requires fitting both source positions and redshifts in all four triplets. \em For the source position and redshift data to lead to the same singlet-pair structure in each triplet that rotates smoothly about the rotation axis in the same direction that the galaxy rotates, and which is at the same time correlated with triplet size, has to be quite remarkable \em. The fact that each triplet's center of mass falls along the rotation axis of the central object is equally remarkable. 

The simple assumption in Fig. 1, that the groups of sources lying approximately equidistant from NGC 1068 have been ejected from the galaxy in separate events, has thus lead to a detailed ejection model with the following characteristics:

a) each event ejects three compact objects (a triplet) along the rotation axis (in either the jet or counter-jet direction) from a point near the center of NGC 1068. At least four separate ejection events have occurred. 

b) immediately after ejection each triplet splits into a singlet and a pair that separate in opposite directions approximately perpendicular to the rotation axis but at different angles to the l-o-s (it is assumed that they separate because the central object is no longer present to contain them in orbit). The position angle given by the line joining the singlet and pair mid-point defines both (i), the position angle of the triplet and (ii), the direction of singlet-pair separation.

c) in each triplet, the pair and singlet are assumed to have similar masses because they lie at equal distances from the rotation axis. (It seems unlikely that random masses and ejection energies would combine fortuitously in each case to produce equal spacings, and more likely that the singlet and pair masses are similar in each triplet).

d) the midpoint between the singlet and pair represents the center of mass of the triplet. In each triplet, the angular offset of its center of mass from NGC 1068, (which always falls along the rotation axis), represents the triplet's motion perpendicular to the l-o-s since its ejection,

e) in addition to their mean redshift components (z$_{\rm mean})$, the measured redshifts of the paired sources (z$_{\rm obs}$) must have equal red- and blue-shifted components $\pm$(z$_{\rm r-b})$, where (1+z$_{\rm obs}$) = ($1\pm$z$_{\rm r-b})(1+z_{\rm mean})$. These are assumed here to be Doppler related and possibly due to residual orbital velocities, implying that each pair itself was initially an orbiting binary. [Note that the observed redshift is also expected to contain a component due to the systemic velocity of NGC 1068 but because this is small (z$_{\rm sys}$ = 0.0038), it has been omitted.]

\section{Discussion}

In Table 2 the sources making up each triplet are listed in column 1, the mean redshift of each pair (z$_{\rm mean}$) in column 2, and the red and blueshifts  (z$_{\rm r-b}$) of the paired sources relative to z$_{\rm mean}$ in col 3. The angular distances of the singlets from NGC 1068 are listed in column 4, and the mean angular distances of the sources making up each pair and triplet are listed in columns 5 and 6 respectively.

\begin{deluxetable}{cccccc}
\tabletypesize{\scriptsize}
\tablecaption{Triplet sources, pair redshifts, and source distances from NGC 1068. \label{tbl-2}}
\tablewidth{0pt}
\tablehead{
\colhead{Triplet(sources)} & \colhead{(z$_{\rm mean})$} & \colhead{(z$_{\rm r-b}$)} & \colhead{Singl. Dist.\tablenotemark{a}} & \colhead{Pair Dist.}  & \colhead{Tripl Dist.} 
}

\startdata
A(1-3,14)  & 0.494  & $\pm$.156  &   43.9  &   48.9  & 46.4 \\  
B(2-5,13)  & 0.761  & $\pm$.166  &  30.4   &   35.1  & 32.7 \\
C(4-6,12)  & 1.101  & $\pm$.215  &  12.1   &   14.4  & 13.3 \\
D(8-11,10) & 1.202  & $\pm$.371  &  6.0    &    7.3  & 6.65 \\
\\
\enddata
\tablenotetext{a}{All distances represent angular distance on the sky from NGC 1068 in min. of arc.}
\end{deluxetable}

a) The central object.

A comparison of the singlet angular distances from NGC 1068 to the mean angular distances of their respective pairs (column 4/column 5) reveals that this ratio remains remarkably constant at 0.86$\pm0.04$ as the triplet size increases. This is shown more clearly in Fig. 5 where the angular distance of each singlet is plotted vs the mean angular distance of its respective pair. This indicates that the singlets and pairs defined in Fig. 1 are tied tightly to a point on the sky aligned closely with NGC 1068, and again suggests strongly that each triplet needs to be considered as a unit. This also shows that the point of origin of the triplets is closely aligned with NGC 1068. When this is taken together with the additional evidence a), that the triplets are ejected along the rotation axis of NGC 1068 and b) that the mean pair redshifts decrease with increasing distance, as found in other similar clusterings of compact sources near active galaxies \citep{arp97a,arp97b,arp98,bur95,bur97a,bur97b,bur96,chu98,pie94,rad97}, it suggests that NGC 1068 is a likely candidate for the central source. 

b) Rotation of the triplet ejection plane.

The position angles, $\beta$, of the triplets on the sky show a continuous rotation from $-101\arcdeg$, through $-84\arcdeg$, $-67\arcdeg$ to $-17\arcdeg$ for triplets A-B-C-D respectively. Note that these values were measured directly from Fig. 3. In order to calculate the corresponding rotation angles $\theta$ (defined as the angular rotation from the l-o-s counterclockwise about the rotation axis, when viewed along the rotation axis towards the galaxy) the rotation axis tip angle $\gamma$ must be known (where $\gamma$ is defined as the amount the rotation axis is tipped toward the observer at the top in Fig. 3). The above values for $\beta$ give the correct values for $\theta$ only when the rotation axis is aligned with the l-o-s ($\gamma$ = 90\arcdeg). This is unlikely to be the case, however. For a more realistic tip angle of $\gamma$ = 18$\arcdeg$, the corresponding values are $\theta$ = -122$\arcdeg$, -71$\arcdeg$, -38$\arcdeg$, and -4.4$\arcdeg$ for the A, B, C, and D triplets respectively. These values have been calculated using the relation tan$\theta_{2} = (1/\alpha$)(tan$\theta_{1})$, where $\alpha, \theta_{1}$ and $\theta_{2}$ are defined as in Fig. 6. Here $\theta_{2}$ is the true rotation angle and $\theta_{1}$ is the apparent angle measured when the rotation axis lies at some angle $\gamma$ to the plane of the sky. The parameters in Fig. 6 are related to $\theta, \beta$ and $\gamma$ as follows: $\alpha$ = sin$\gamma$ (for a circle with a radius 1);  $\theta = (\theta_{2} + 270\arcdeg); \beta = (\theta_{1} + 270\arcdeg$).  

\begin{figure}
\hspace{-1.5cm}
\vspace{-2.0cm}
\epsscale{1.1}
\plotone{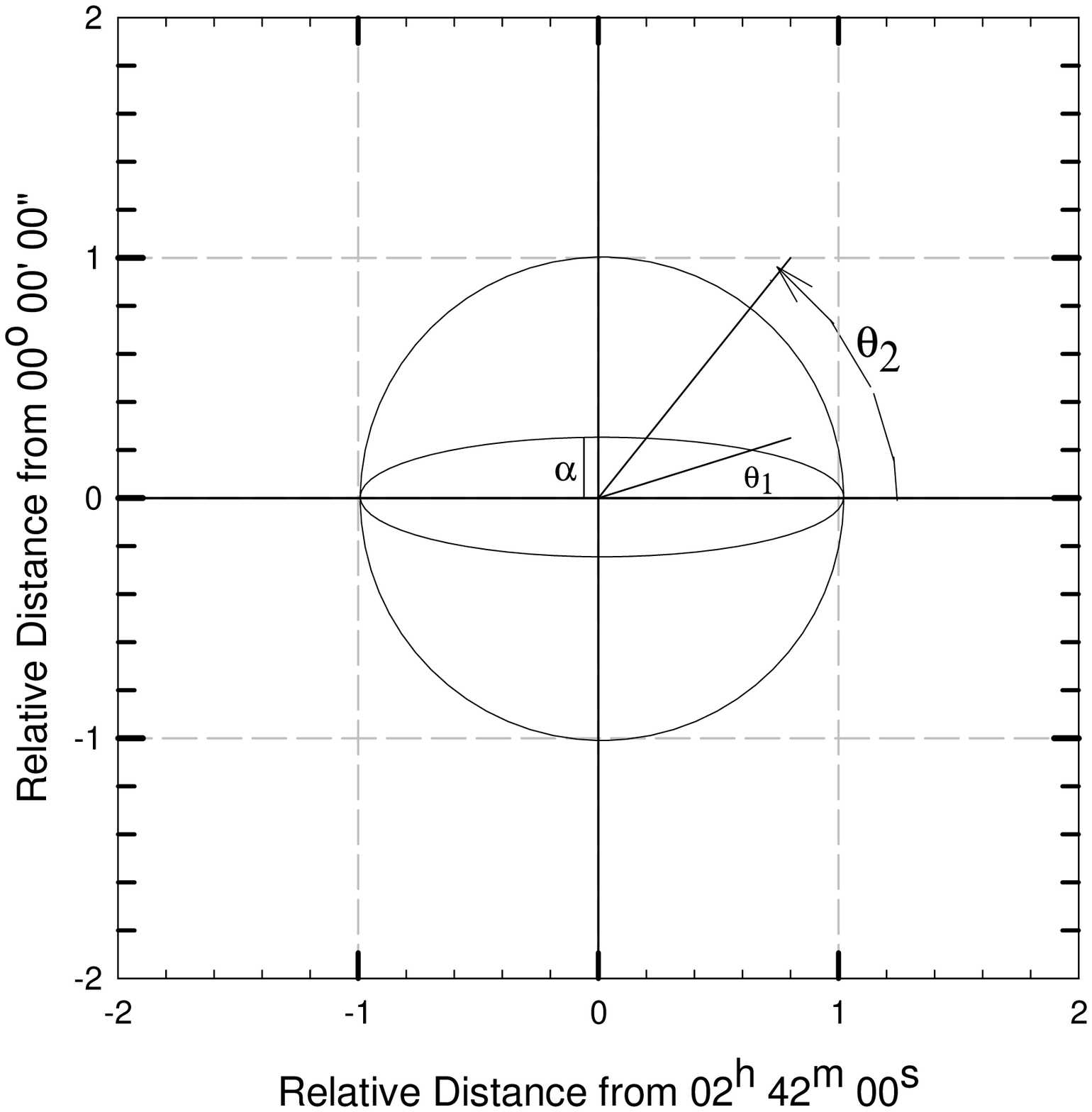}
\caption{\scriptsize{Diagram used to convert projected angles ($\theta_{1}$) to actual rotation angles ($\theta_{2}$) as described in the test.\label{fig6}}}
\end{figure}

\begin{figure}
\hspace{-1.5cm}
\vspace{-1.5cm}
\epsscale{1.05}
\plotone{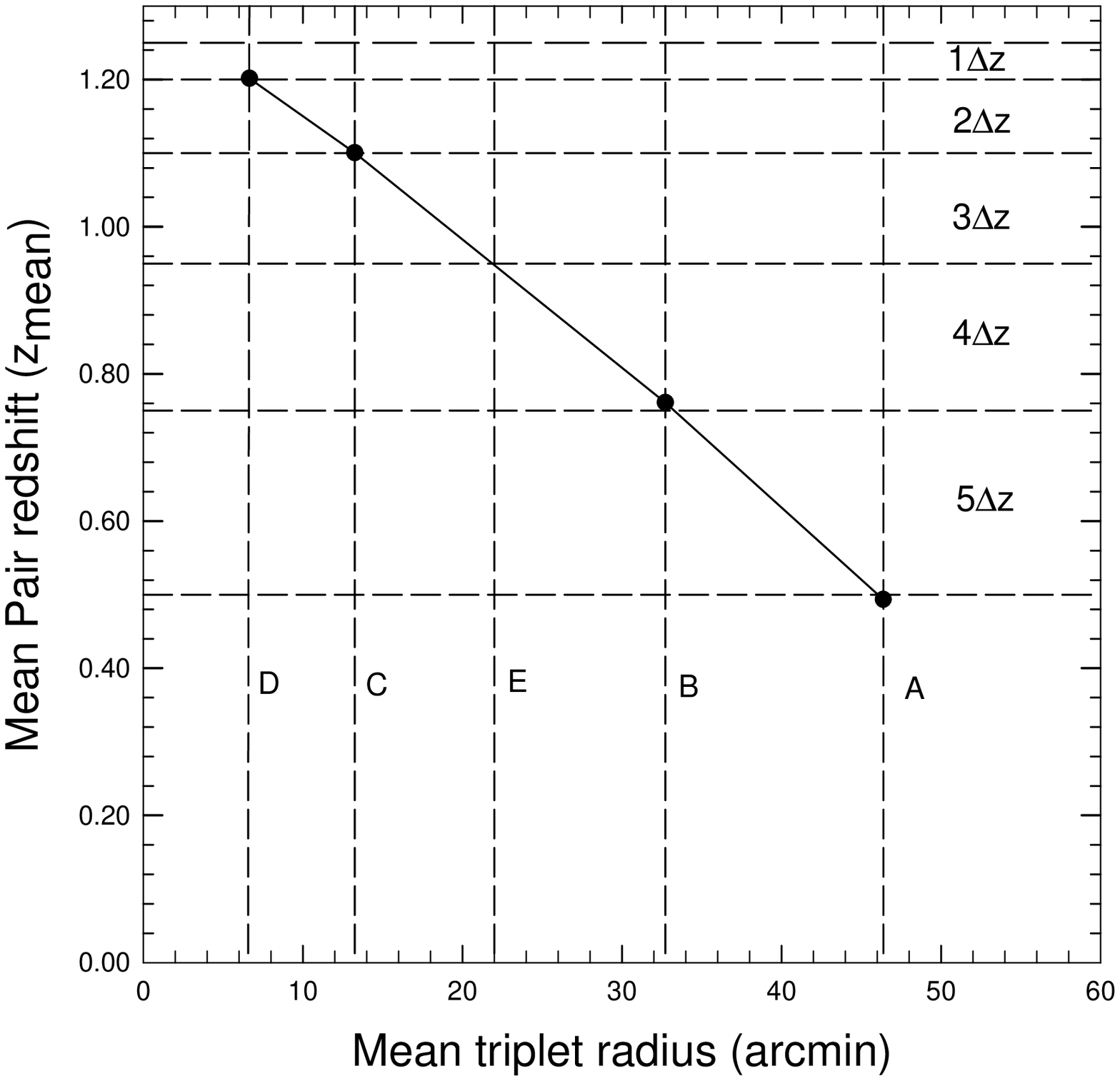}
\caption{\scriptsize{Mean pair redshifts plotted vs mean angular distance from NGC 1068 of sources making up each triplet. The horizontal dashed lines indicate the position of perfectly quantized redshifts with a quantization interval $\Delta z = 0.05$. The mean distance predicted for a possible missing triplet is labeled E. \label{fig7}}}
\end{figure}

As noted earlier, if it is assumed that the larger triplets were ejected first, rotation is in the same sense as found for the galaxy and central torus, with gas and water vapour maser emission redshifted in the West and blueshifted in the East \citep{gre96}. 

c) Pair-singlet separation velocities, event ages and galaxy rotation rate.

If inside each triplet the non-Doppler components of the singlet and mean pair redshifts are identical [which might be expected for similar masses if the intrinsic component is gravitationally produced, or if mass and intrinsic redshift are age related \citep{nar80,bel01}, the differences between the singlet and mean-pair redshifts will give the l-o-s component of the pair-singlet separation velocity. Since the redshifts for all sources in triplets A and B have been measured, it is possible to test this. Since the rotation angle of triplet A is $>-90\arcdeg (-122\arcdeg$ for $\gamma = 18\arcdeg$), the pair in triplet A must be blueshifted relative to singlet A. For triplet B, with a position angle $<-90\arcdeg (-71\arcdeg$ for $\gamma = 18\arcdeg)$, the opposite must be true. From cols 2 and 3 of Table 3 this can be seen to hold true, indicating that the assumption of equal intrinsic redshift components may be a valid one, allowing the l-o-s singlet-pair separation velocities to be estimated directly from the difference between the singlet and mean-pair redshifts.

\begin{deluxetable}{cccccccccc}
\tabletypesize{\scriptsize}
\tablecaption{Ejection redshifts and velocities. \label{tbl-3}}
\tablewidth{0pt}
\tablehead{
\colhead{Triplet} & \colhead{z$_{\rm obs}$(sgl)} & \colhead{(z$_{\rm mean}$)} 
& \colhead{(z\tablenotemark{a}$_{\rm separat.}$)}  
& \colhead{vel\tablenotemark{b}$_{\rm separat.}$}  
& \colhead{Angl.\tablenotemark{c}}  
& \colhead{z\tablenotemark{d}$_{\rm ejection}$} 
& \colhead{vel\tablenotemark{e}$_{\rm ejection}$} 
&  \colhead{v\tablenotemark{f}$_{\rm perpend}$}  
&  \colhead{Ang. Disp.\tablenotemark{g}}\\
 
  &  &  &  &  \colhead{(km.s$^{-1}$)} & \colhead{(deg)} &  &   \colhead{(km.s$^{-1}$)} 
&  \colhead{(km.s$^{-1}$)} & \colhead{($\arcmin$)}  \\
  &  &  &  &  \colhead{($\times10^{4}$)} &  &  &  \colhead{($\times10^{4}$)} & \colhead{($\times10^{4}$)}
}
\startdata
A &0.655 &0.494 &$\mp$0.054  &$\mp1.58$  &-122  &0.101  &3.03  &2.57  &39.0 \\  
B &0.684 &0.761 &$\pm$0.0233 &$\pm0.69$  &-71   &0.071  &2.13  &2.01  &22.3 \\

\enddata


\tablenotetext{a}{Pair and singlet separation redshifts calculated as descibed in the text} 
\tablenotetext{b}{Component of pair and singlet separation velocities parallel to the l-o-s}
\tablenotetext{c}{Rotation angle $\theta$ for $\gamma = 18\arcdeg$ measured counter-clockwise from l-o-s}
\tablenotetext{d}{Radial ejection redshift (col 4/cos$\theta$)}
\tablenotetext{e}{Radial ejection velocity (from triplet center of mass)}
\tablenotetext{f}{Velocity component perpendicular to the l-o-s [(col 8)$\times$sin$\theta$]}
\tablenotetext{g}{Mean angular displacement of singlet and pair midpoint from triplet center of mass}
\end{deluxetable}

Making this assumption it is now possible to determine the pair-singlet separation velocities and use them to calulate event ages. In Table 3 the separation redshift of the pairs and singlets in triplets A and B are listed in col 4. These are produced by the l-o-s Doppler motion of the pairs and singlets as they move out from their respective centers-of-mass. They have been calculated from the following relations: 

\scriptsize{(1 + z$_{\rm obs}$)$_{\rm singl}$ = (1 + z$_{\rm sys}$)(1 + z$_{\rm separat}$)$_{\rm singl.}$(1 + z$_{\rm t}$)(1 + z$_{\rm int}$)$_{\rm singl}$}
\normalsize 

and

\scriptsize{(1 + z$_{\rm mean}$) = (1 + z$_{\rm sys}$)(1 + z$_{\rm separat}$)$_{\rm pair}$(1 + z$_{\rm t}$)(1 + z$_{\rm int}$)$_{\rm pair}$}

\normalsize
Here z$_{\rm t}$ is the redshift of the center of mass of the triplet relative to the galaxy and (z$_{\rm int}$)$_{singl}$ is the intrinsic component of the singlet, assumed to be equal to the intrinsic component of the pair (z$_{\rm int}$)$_{\rm pair}$. Since (z$_{\rm separat}$)$_{\rm pair}$ is assumed to be equal to (-z$_{\rm separat}$)$_{\rm singl}$, and z$_{\rm sys}$ and z$_{\rm t}$, are common to the two equations, the only unknown is the separation velocity which can then be solved for. This must be divided by cos$\gamma$ to take account of the tip angle of the rotation axis. The resulting values are listed in col 4. The ejection redshifts in col 4 are converted to l-o-s velocities in col 5 of Table 3 using the non-relativistic relation v = cz. The radial separation velocity of the singlets and pairs are listed in col 8. Col 9 gives the velocity perpendicular to the l-o-s and col 10 the mean angular displacement on the sky of the singlets and pair midpoints from each triplet's center of mass. 

If the compact objects are located near NGC 1068, the values found for the ejection velocity perpendicular to the l-o-s (col 9), along with the angular displacement on the sky (col 10) give the elapsed time since each event occurred and therefore give the approximate age of each triplet. At the distance of NGC 1068 1$\arcmin$ = 4.2 kpc. These numbers thus result in ages near $6.08\times10^{6}$ and $4.46\times10^{6}$ yrs for the A and B triplets respectively (for a tip angle of $\gamma$ = 18$\arcdeg$). Thus triplet B is younger than triplet A in agreement with the assumptions that larger triplets are older, and that the rotation sequence is from A to B. 

Between the A and B events the position angle (given by $\theta$) rotates through $\sim51\arcdeg$. Since this amount of rotation takes 1.62$\times10^{6}$ yrs, one complete revolution will take  $\sim1.1\times10^{7}$ yrs, or a factor of 25 shorter than the time it takes the Sun to make 1 revolution in our own Galaxy. This value is in good agreement with the value (6$\times10^{6}$ yrs) calculated for the rotation period of the nucleus of NGC 1068 using recent kinematic results \citep{all01}, when it is kept in mind that it applies for $\gamma = 18\arcdeg$ only, and this angle is not accurately known. Furthermore, there may be some evidence that this angle changes over $10^{7}$ yrs (see below). This result is interpreted as indicating that the changes in triplet rotation angle are tied closely to the rotation of the galaxy.  

The above numbers give an angular rotation rate of 3.17$\times10^{4}$ yrs/deg. The rotation angles of triplets C and D thus lead to ages for these triplets of 3.42$\times10^{6}$ yrs and 2.81$\times10^{6}$ yrs respectively, for $\gamma = 18\arcdeg$.

d) Quantization in z$_{\rm mean}$.

Examination of the z$_{\rm mean}$ components [where z$_{\rm mean}$ = (z$_{1}+$z$_{2})/2$] reveals that not only do they fall off smoothly with increasing distance from NGC 1068, they do so in a regular, harmonically related or "quantized", manner. The quantization is in units of $\Delta$z = 0.05, decreasing outwards, as the triplet size increases, (in redshift steps of z = 1$\Delta$z, 2$\Delta$z, 3$\Delta$z,...) from a maximum redshift of z$_{\rm mean}$ = 1.25 near the central point, to a minimum of z$_{\rm mean}$ = 0.5 in the most remote triplet. 
This can be seen more clearly in Fig. 7, where the z$_{\rm mean}$ values have been plotted (filled circles) versus the mean angular displacement of the sources in each triplet from NGC 1068 (col 6 in Table 2). The locations of perfectly quantized redshifts are indicated by the horizontal dashed lines and show good agreement with the measured values. The "quantized" redshifts are simply the mean of two measured values. Also implicit in this figure is the indication that there is a missing triplet with z$_{\rm mean}$ redshift near 0.95. Although its redshift has not been measured, source 9 is located at approximately the correct angular distance ($22\farcm6$) from NGC 1068 to be part of such a triplet.

Since the singlet and pair midpoint in each triplet are separating from one another at different angles to the l-o-s (given by $\beta$), the mean redshifts of the 4 pairs would be expected to have superimposed Doppler components. It is thus not clear yet whether the quantization is a real quantization in an intrinsic redshift component, or simply due to some fortuitous combination of Doppler and intrinsic redshifts. Obtaining redshifts for sources 10 and 12 should help to clarify this. In Fig. 7, as the redshifts decrease in what appear to be quantized steps, the distance to the relevant triplets increases in a similar manner such that the relation between increasing angular displacement and mean-pair redshift varies smoothly and approximately linearily. A similar quasi-linear relation was found previously for sources near NGC 3516 \citep{chu98}. Note that this quantization appears in only one component (z$_{\rm mean}$) of the measured redshifts. Since the systemic redshift of NGC 1068 is known, the quantized components cannot contain a large unknown cosmological component. This apparent quantization is therefore unlikely to be related to the periodicity reported in other studies \citep{bur01,kar77} where the entire measured redshifts have been used.

e) Is there a fifth triplet?

The manner in which the singlet redshifts vary with distance from NGC 1068 cannot be determined yet because the redshifts of singlets 10 and 12 have not been measured. In Fig. 7 the missing triplet postulated above has been labeled E and is predicted to be located at an angular distance of $\sim22\arcmin$ from NGC 1068. This information, and position information from the other four triplets, is being examined in an attempt to pinpoint on the sky the most likely locations of the two sources required to complete triplet E (Bell, 2001, in preparation). 

f) Proper motions of the ejected sources.

The age of triplet A can be used together with the mean angular displacement of its sources from NGC 1068 (Table 2, col 6) to calculate the proper motion. This is found to be $<0.7$ milliarcsec per yr which would be difficult to detect even if observations were spaced over many years.

g) Progenitor Configuration.

This paper has examined the position and redshift data of the compact objects near NGC 1068 as they currently exist and no attempt has been made to determine their configuration prior to ejection. Although it may be premature to speculate on what the progenitor source configuration might have been, there are some things that are worth noting. In Fig. 4 the position angle of the paired sources ($\omega)$ can be seen to rotate smoothly from A to D, ending up approximately perpendicular to the present position of the rotation axis and aligned with the major axis of the central torus. If this change is produced by a simultaneous change in the tip of the rotation axis with time the present position angles of the pairs are at least consistent with the possibility that they all originally orbited in the major axis plane of the galaxy.

Since the paired sources have equal-and-opposite spectral shifts (z$_{\rm r-b}$), it is argued here that prior to ejection each pair was an orbiting binary. Note that the z$_{\rm r-b}$ values (as high as z = 0.37 in pair D) are too large to be explained as components of the modest ejection velocities and seem more likely to be residual orbital velocities. Also, in Fig. 4 the letters "r" or "b" next to the paired sources indicates whether the source is redshifted or blueshifted relative to the mean pair redshift. The blueshifted pair members (1,2,4,8) all lie west of the redshifted members (3,5,6,11), and any proposed progenitor configuration must be able to explain this. It is of interest to note that if the displacements of the pair midpoints from the galaxy are removed, all the blueshifted objects fall on the West of the galaxy, and the redshifted ones on the East. This seems to imply that if the pairs were initially in orbit, they all orbited in the same direction.

h) Hidden blueshifts.

It is worth pointing out that the above ejection model demonstrates how large blueshifted components [negative z($_{\rm r-b}$) values] can be present but still be completely camouflaged by even larger intrinsic redshifts.

i) Redshifts of sources 10 and 12.

Until the redshifts of sources 10 and 12 can be measured it will not be possible to determine the size of the intrinsic component of z$_{\rm mean}$, since this redshift component will also contain a component due to the N-S, l-o-s triplet ejection velocity (if $\gamma \neq 0$). However, that component can be estimated to be at least as high as z = 0.6 (for $\gamma > 0$). These results do allow the redshift of source 12 to be roughly estimated for a future test of the model. Since sources 10 and 12 are both moving toward us in the model, their redshifts must be less than the z$_{\rm mean}$ values of their respective pairs. If the intrinsic redshift component of z$_{\rm mean}$ is equal to that of its associated singlet, as assumed above, then the redshifts of 10 and 12 should differ from z = 1.2 and 1.1, respectively, by twice their l-o-s ejection velocities. This difference will be small if the radial ejection velocities are similar to those found for the singlets in triplets A and B. The redshift of source 12 is estimated here to be z = 0.63$\pm0.1$ using the estimated triplet age and the angular separation of the singlet from the rotation axis. No attempt has been made to estimate a value for the redshift of source 10 because the angular distances involved are too small to estimate accurately. 

\section{Conclusion}

If the four triplets defined by the boxes in Fig. 1, whose component sources all lie at a similar distance from NGC 1068, have been ejected from the galaxy in separate events, it has been shown that the present positions of the sources on the sky can lead to a detailed, reasonable, internally consistent ejection model. For the source position and redshift data to lead to the same singlet-pair structure in each of the four triplets, that rotates smoothly about the rotation axis in the same direction that the galaxy rotates, and which is, at the same time, correlated with the triplet size and age, is remarkable. For the triplet ejection direction to lie along the rotation axis of the galaxy, the one direction most closely associated with the ejection of matter from and active galaxy, is equally remarkable. This suggests strongly that the objects have been ejected from NGC 1068. However, the modest ejection velocities found indicate that a significant component of the measured redshifts must be intrinsic. The consequences of this conclusion are enormous. 

\section{Acknowledgements}
I wish to thank an anonymous referee for several helpful suggestions pertaining to the presentation of these results.

\end{document}